\documentstyle[titlepage,fleqn,12pt]{article}
\begin{document}
\title{ Optical diffraction from fractals \\
with a structural transition }
\author{
\\
\\
Felipe P{\'e}rez-Rodr{\'{\i}}guez \\
{\it Instituto de F{\'{\i}}sica, Universidad Aut{\'o}noma de Puebla,}\\
{\it Apdo.Post. J-48, Puebla 72570, Pue., M{\'e}xico}
\\
\\
Wei Wang \\
{\it Physics Department, Nanjing University, People's Republic of China}
\\
\\
Enrique Canessa\\
{\it ICTP-International Centre for Theoretical Physics, Trieste, Italy}
}
\date{}
\newpage
\maketitle
{\baselineskip=25pt
\begin{abstract}

A macroscopic characterization of fractals showing up a
structural transition from dense to multibranched growth
is made using optical diffraction theory.  Such fractals
are generated via the numerical solution of the 2D Poisson
and Biharmonic equations and are compared to more 'regular'
irreversible clusters such as diffusion limited and Laplacian
aggregates.  The optical diffraction method enables to identify a
decrease of the fractal dimension above the structural point.

\vskip 2cm

\end{abstract}
}
\baselineskip=25pt
\parskip=0pt

Fractal surfaces observed in nature can become very complex due to
structural transitions that are generated during their
growing processes.  Examples of such phenomena can be found
on bacterial colonies \cite{Ben92} and electrochemical deposition
experiments \cite{Fle92}.  The observed fractal patterns exhibit
a similar class of complex structure despite the fact that the
mechanisms involved are clearly different.  It has been found
that, at a certain threshold distance, these systems display an
intriguing structural transition from dense to multibranched growth
behaviour of which not much is yet understood.

In the current literature, two different approaches attempting to explain
and reproduce a structural transition during irreversible growth have
been proposed.  In the first one \cite{Lou92,Cas93,Wan93a} the transition
has been derived by solving the Poisson equation on a squared lattice
-that becomes dependent on the potentials at two boundaries, the distance
between them, and a screening length.  Hence, it has been argued that
screening, due to free charges, strongly diversifies the patterns that
grow in the presence of electrostatic fields \cite{Nie84}.

An extention of this problem is the second approach due to the authors
\cite{Wan93b,Can93}, which is based on a Biharmonic equation
in two-dimensional (2D) isotropic defect-free media. By discretizating
the Biharmonic equation we have proved that a transition from dense to
multibranched growth can also be a consequence of a different
coupling of displacements during the pattern formation.  Within the
Biharmonic model the transition appears when the growth velocity at the
surfaces present a minimum as also occurs within the Poisson growth.  The
later implies that both approaches, {\em i.e.} Poisson and Biharmonic
growth, describe a similar class of complex structural transition phenonema
from two different perspectives and on different systems.

It is the aim of this work to deal with a macroscopic characterization
of fractals showing up a structural transition on growing by using the
optical diffraction method.  The formalism
adopted is not new, as it is an extention of what has been largely
applied to analyse optical diffraction from rather 'regular'
sctructures like diffusion limited aggregates (DLA) \cite{Wit83} and
Laplacian fractals (see, {\em e.g.,} \cite{All86,Kay89,Ber91,Kor92}),
besides Koch clusters \cite{Pei93,Uoz9091}.

We shall expand on these ideas to study the optical diffraction patterns
of fractals displaying a transition from dense to multibranched structure.
Fractal growth is here numerically simulated using the Poisson and Biharmonic
approaches and the results are compared with the well-known cases of DLA and
Laplacian growth.  We investigate the effect of the structural transition
on the diffracted intensity and, consequently, on the fractal dimension
$d_{f}$ for all different types of clusters.

For completeness we shall give a brief description of the optical diffraction
formalism.  Let us start considering a fractal structure composed of $N$
identical and similarly oriented particles on the plane $x-y$.  The position
of their centers of mass is given by ${\bf R}_{n}=(x_{n},y_{n})$, where
$n=1,...,N$. In this system we investigate the Fraunhofer diffraction
pattern for the fractal, assuming that each particle corresponds to one
aperture. This assumption is experimentally realizable even in
the case of dark clusters, since they can be photographed on
high contrast films which are later lightened \cite{Ber91}.
Supposing that the incident plane wave propagates parallel to the $z$-axis,
the diffraction amplitude due to $N$ 'apertures'
can be expressed as \cite{Bor75}
\begin{equation}\label{eq:a1}
A({\bf k})= C \int _{S} e^{-i{\bf k}{\bf r}}d^{2}{\bf r}
	\sum _{n=1}^{N}e^{-i{\bf k}{\bf R}_{n}} \;\;\; ,
\end{equation}
where $C$ is a constant factor, $S$ denotes the surface occupied by
one particle. The vector {\bf k} in Eq.(\ref{eq:a1}) is the
component, parallel to the $xy$-plane, of the scattered wave vector.
Its modulus is given by $k = \frac {2\pi }{\lambda }
{\rm sin}\theta \approx \frac {2\pi }{\lambda } \theta$, where
$\theta $ is the angle (small in practice) which the scattered
wave vector makes with the $z$-axis and $\lambda $ is the wave length
of the incident light.

The form factor $F({\bf k})$, corresponding to the
intensity scattered by one 'aperture', is determined by the integral
in Eq.(\ref{eq:a1}). The sum over $n$ in Eq.(\ref{eq:a1}) determines the
structure factor
\begin{equation}\label{eq:a4}
S({\bf k})\equiv \mid \frac {1}{N} \sum _{n=1}^{N}
e^{-i{\bf k}{\bf R}_{n}}\mid ^{2} \;\;\; .
\end{equation}
In the case that $k$ is smaller than $a^{-1}$ ($a$ being the size of an
elementary particle), the form factor is practically constant, ({\em i.e.},
$F({\bf k})\approx 1$) and the light distribution in the diffraction
pattern is given by the structure factor such that the normalized
diffraction intensity $I({\bf k})\approx S({\bf k})$, with $ka<1$.

The factor S({\bf k}) of Eq.(\ref{eq:a4}) depends on the distribution of
particles in the cluster and is related to its fractal dimension. This
relation can be found by using density-density correlation
functions (see, {\em e.g.,} \cite{Wit83,Kor92} and references
therein).  Accordingly, for $L^{-1}<k<a^{-1}$ ($L$ is the size of the whole
aggregate) the expected value of the intensity -or, alternatively,
of the structure factor $S({\bf k})$- is
\begin{equation}\label{eq:lq2}
<I({\bf k})>= \int d^{2}{\bf R} e^{-i{\bf k}{\bf R}}
<\rho({\bf R}_{0}) \rho ({\bf R}+{\bf R}_{0})> \;\;\; ,
\end{equation}
where
\begin{equation}
<\rho({\bf R}_{0}) \rho ({\bf R}+{\bf R}_{0})>=
\int d^{2}{\bf R}_{0} \rho ({\bf R}_{0})
\rho ({\bf R}+{\bf R}_{0}) \; \sim R^{-\alpha}\;\;\; ,
\end{equation}
represents the density-density correlation function.
For fractal aggregates this relation obeys a power law variation \cite{Wit83},
where the exponent $\alpha$ is related to the Hausdorff dimension
\cite{Man77} $d_{f}=d-\alpha$ with $d$ the Euclidean space dimension. This
variation of the density-density correlation function leads to the
power law behavior of the diffracted intensity as a function of the wave
vector $I({\bf k})\sim k^{-d_{f}}$ \cite{Wit83}.
Thus, it is possible to estimate the fractal dimension $d_{f}$ from
diffraction patterns.  In particular we shall see that this is valid for
fractal structures displaying a structural transition as those at hand.

In the following we shall calculate the quantities $S({\bf k})$ and
$<I({\bf k})>$ of the diffraction intensity employing Eqs.(\ref{eq:a4})
and (\ref{eq:lq2}) and shall search for a power law behaviour.
For simplicity we shall assume that the apertures (or particles) have a
squared form $a\times a$ with a distribution obtained by numerically
simulating fractal growth.
In the simulations the particles are distributed on a squared lattice, and
the centre-of-mass coordinates $x_{n}$, $y_{n}$ are assumed to take the
following discrete values
\begin{eqnarray}\label{eq:a7}
x_{n}= m_{n}a \; , \;\; m_{n}= 0,\pm 1, \pm 2, ...  ; & \;\;\;\;
y_{n}= \ell_{n}a \; , \; \; \ell_{n}= 0,\pm 1, \pm 2, ...  ,
\end{eqnarray}
Therefore, according to Eqs.(\ref{eq:a4}) and (\ref{eq:a7}),
the structure factor is calculated by the formula
\begin{eqnarray}\label{eq:a10}
S(k_{x},k_{y})& \equiv & \mid  \frac {1}{N}
\sum _{n=1}^{N} e^{-ia(k_{x}m_{n}+k_{y}\ell_{n})}
\mid ^{2} \;\;\; .
\end{eqnarray}

Following the DLA simulation model \cite{Wit83}, we add particles to the
growing clusters one at the time undergoing a random walk that starts from a
point on a variable circle centered on the simulation box.  In this process
the particles are deposited adjacent to occupied lattice sites to then
start off again the random walk of a new particle at another randomly chosen
position, and so on. An example of DLA is shown in Fig.1 (upper left-corner).

Laplacian fractals are generated using the growth law
(the so-called dielectric breakdown model) \cite{Nie84}
\begin{equation}\label{eq:w1} \nabla^{2}\phi =0 \;\;\; .
\end{equation}
We use lattice sites enclosed within a circle of (normalized) radius
$r=\sqrt{i^{2}+j^{2}}=100$ such that $\phi^{o}$ and $\phi^{i}$ are unity
and zero at the outer circular boundary and the inner growing aggregate,
respectively.  Seed particles are placed centered in the simulation box.
The discretazion procedure then follows standard techniques \cite{Nie84}
till solutions of the discretization of Eq.(\ref{eq:w1}) converge to a
desidered accuracity.  The stochastic growth probability $P$ adopted
(at the grid site $(i,j)$) is assumed to be proportional to the
local field \cite{Nie84}, {\em i.e.}
\begin{equation}\label{eq:w2}
P_{ij}= \frac{\mid \phi_{i,j}\mid^{\eta} }{\sum \mid
	\phi_{ij}\mid^{\eta} } \;\;\; ,
\end{equation}
where the sum runs over nearest neighbor sites to a cluster.  Herein we set
$\eta =1$ for the sake of simplicity.  An example of Laplacian fractal is
shown in Fig.1 (upper right-corner).

The effect of screening on fractal structures growing under electrostatic
fields was reported in Ref.\cite{Lou92,Cas93}.  Within this model the
previuos Laplacian equation is replaced by the linearized
Poisson-Boltzmann equation \begin{equation}\label{eq:p1}
\nabla ^{2}\phi =\lambda ^{2}\phi \;\;\; .
\end{equation}
The origin of screening lies in the presence of free charges and leads to a
rich variety of patterns -see the example in Fig.1 (lower left-corner).  The
model  introduces a new length scale, {\em i.e.},
$\lambda$, and a nontrivial dependence on the boundary conditions which is
responsible for a structural transition on growing.  The patterns can have a
fractal character at scales shorter than $\lambda$, be Eden-like, or
grow dense, to then follow the transition from dense to single branch
growth, which is characterized by a change in the sign of the electrostatic
field at the aggregate \cite{Lou92}.  However, one of us \cite{Wan93a}
has demostrated that the transition found using Eq.(\ref{eq:p1})
is altered by the existence of a critical field in the growth of the pattern.

On the other hand, the alternative Biharmonic model under consideration is
based on the discretization (in a square lattice of size $L\times L$) of the
Biharmonic equation \cite{Wan93b,Can93}
\begin{equation}\label{eq:ww1}
\nabla^{2}(\nabla^{2} u)=0   \;\;\; .
\end{equation}
A crucial difference with respect to the above Laplacian and Poisson
models, is that within the Biharmonic model, iterative procedures are carried
out around {\em thirteen} next nearest neighbours -and not on {\em four} as
for Eqs.(\ref{eq:w1}) and (\ref{eq:p1})-
on equal grounds.  Because of this the formation of connected patterns
within the Biharmonic equation becomes non trivial, hence this model is
more involved than Laplacian or Poisson growth.  To generate Biharmonic
fractal patterns -as seen in Fig.1 (lower right-corner)- we set for
simplicity the derivative  boundary condition,
that is necessary along the radial-direction, equal to zero and the growth
probability $P$ proportional to $\nabla^{2} u$, (corresponding to the
potential in Eq.(\ref{eq:w2})).

We focus now on the results of diffraction theory. Due to the complexity
of the present numerical calculations for growing fractals (8 hrs circa of
CPU Convex time for each run), our results are based in a statistics of
only few different clusters for each class of complex structures
considered.  The data shown next is displayed for illustrative purposes,
whereas the straight lines (used to estimate $d_{f}$) are the result of our
crude, but representative, statistics.

To this end we can add that the accuracity achievable depends on the
wavevector regime considered.  Variations of $<I(k)>$ decrease on
increasing $k$ and, accordingly, the error also decreases.
The reason for this essentially lies on the number of particles considered
(which form the fractal) and also on the shape of the box used in
simulations (squared in the present case).
For $k$ values above the transition point $(\sim 1/0.6L)$ we estimate the
error of our calculations to be less than $1\%$, whereas for $k$ values
near to the system size $\sim 1/L$ the errors increase to about $5\%$. Thus,
to make a better estimation of $d_{f}$ it will be necessary to obtain
statistic for $<I(k)>$ over a wide range of $k$-values.

As an example we plot in Fig.2 the diffracted intensity $I$ as
a function of the wave vector ${\bf k}$ at the plane $k_{y}=0$.
The diffraction pattern corresponds to a Biharmonic fractal with
1500 particles (each of size $a$) similar to the one displayed in Fig.1.
This plot shows strong fluctuations associated to the fractal structure of
the aggregate. It can be seen that fluctuations between zeros of the form
factor ($2\pi n \leq \mid k_{x} \mid a \leq 2\pi (n+1)$; $ n=1,2,...$)
are of the same type. That is, the structure fractor
turns out to be a periodic function of the wave vector:
\begin{equation}
S(k_{x} +2\pi m/a) = S(k_{x}) \ \ ;
\ \ m=0,\pm 1,\pm 2,...
\end{equation}
This fact is a direct consequence of the assumed distribution
of particles on a squared lattice ({\em c.f.}, Eq.(\ref{eq:a10})).
When the magnitude of the wave vector is sufficiently large, namely
$k_{x}a \gg 1$, the intensity decreases considerably due to the form factor
$F(k_{x}) \sim k_{x}^{-2}$.

Information about the fractal dimension $d_{f}$ may be obtained at
low values of the wave vector ( $L^{-1} \ll \mid k_{x} \mid \ll a^{-1}$).
Therein $F(k_{x})\approx 1$ and $I(k_{x})\approx S(k_{x})$.
It is noteworthy that in the case of deterministic fractals \cite{All86}
the method, used to obtain the fractal dimension $d_{f}$, consists
of averaging the structure factor over each of its frequency bands,
which are scale invariant. The so averaged structure factor $<S(k_{x})>$
varies according the power law $k_{x}^{-d_{f}}$. In our example the structure
factor for a random Biharmonic fractal, as shown in Fig.3, also decreases at
$k_{x}a \ll 1$, but it has no scale-invariant frequency bands. Therefore,
we cannot apply the preceeding relation between $S(k_{x})$ and $k_{x}$ for
random fractals. In order to avoid this problem, we have calculated the
averaged intensity defined by the expression
\begin{equation}\label{eq:a16}
<I(k)>\equiv \frac {1}{2\pi} \oint d\phi I({\bf k}) \ \ ;
\ \ \ k_{x}=k{\rm cos}\phi, \ \ k_{y}=k{\rm sin}\phi,
\end{equation}
As we shall see next, this quantity has an evident power law behavior
as a function of the modulus ($k$) of the wave vector.

Our numerical results for $<I(k)>$ of diffussion limited and Laplacian
aggregates are given in Fig.4 by full and open squares, respectively.
The averaged intensity $<I(k)>$ is smoother than
$I({\bf k})$ and decreases as $k^{-d_{f}}$.
The fractal dimension $d_{f}$ is obtained by adjusting the log-log plot
of $<I(k)>$ to a straight line as indicated in these figures.  So, the
fractal dimensions for DLA and Laplacian patterns obtained from these
curves are $d_{f}=1.72$ and $d_{f}= 1.70$, respectively. These $d_{f}$
values are closed to the values calculated via the box counting method,
namely, by counting the number of particles $N(r)$ inside an increasing
radius $r$ (around a seed particle) -see, {\em e.g.}, \cite{Can91}.

We shall see next that
a structural transition during fractal growth leads to a change in their
fractal dimension $d_{f}$.  The slopes of the curves in Figs.5 and 6 show
the decrease of the fractal dimension for Poisson and Biharmonic agregates,
respectively.  (These were roughly estimated similarly to Figs.3 and 4).

In the case of Poisson fractals (as the one in Fig.1: lower left-corner),
$d_{f}$ varies from 1.60 -full squares in Fig.5- (by counting N=650 particles
before the transition) to 1.44 as deduced by fitting the open squares to a
second line. Before the structural
transition the $d_{f}$ value for a typical Biharmonic fractal (as shown
in Fig.1: lower right-corner) is equal to 1.66 -full squares in Fig.6- being
quite closed to the corresponding fractal dimension of diffussion limited
agregates and Laplacian fractals in Fig.4, respectively.  After
the transition point the $d_{f}$ obtained from the slope in Fig.6 (open
squares) for  $<I(k)>$ of a Biharmonic pattern reduces to a value of
1.59.

For circular geometry, a very crude analytical analysis of the
Biharmonic Eq.(\ref{eq:ww1}) implies that the transition point for
Biharmonic fractals is approximately located at a distance far from the
center equal to $r_{\ell}/(L/2)\approx e^{-1/2}$ by considering
displacements of the growing surfaces in a continuous limit in reasonable
agreement with numerical simulations \cite{Wan93b,Can93}.
By comparing the structures in Fig.1, it can be seen that Poisson fractals
display a transition point which appears at a rather shorter distance than
irreversible Biharmonic structures (the latter enclosed by a circle).  This
structural transition appears at a point in which the growth velocity of
the active zone of the clusters exhibits a minimum similarly to what occurs
within Poisson growth, but with different magnitudes \cite{Lou92,Wan93c}.
This fact explains the small differences observed in the value of $d_{f}$
after the structural transition within the Poisson and Biharmonic approaches.

Results of optical diffraction thus enables to
identify and relate changes in the fractal dimension $d_{f}$ of
aggregates to variations in the diffracted intensity as a function
of the wave vector.  We have seen that there is indeed a decrease of
$d_{f}$ above the structural transition, which appears beyond one
half of the system size (assuming $\eta =1$ in Eq.(\ref{eq:w2})) in accord
with early crude estimations \cite{Wan93b,Can93,Wan93c} made using the box
counting method, even if it falls beyond the transition point.
Different magnitudes of $d_{f}$ above the structural transitions
are due to the different growth velocities obtained from the
Poisson and Biharmonic approaches.

Our present findings may -in principle- be experimentally confirmed.
The averaged intensity $<I(k)>$ might be determined with an experimental
arrangement as in the one described in \cite{All86}.  Therein, a
photomultiplier is connected to a multichannel analyzer to thus
record $I(k_{x},k_{y})$ and the displacement of the photomultiplier
is controlled by a high-precision motorized micrometer. Then, after
scanning the diffraction patterns, the average of intensity over
concentric circles ($<I(k)>$) is obtained.

Concerning our simulations, it would be interesting to extend them
to values of $\eta \neq 1$.  This is so because by tuning $\eta
\rightarrow 0$, the structural transition now corresponds to a
"dense-to-multibranched transition", whereas for $\eta \rightarrow
\infty$ one obtains a "transition from slow to faster growth" \cite{Wan93c}.

\begin{center}
{\bf  Acknowledgments}
\end{center}

One of the authors, F.P-R., would like to thank the partial support by the
Consejo Nacional de Ciencia y Tecnolog\'{\i}a (CONACyT, M\'exico) under
Grant No.2048-E-9302. The others two, W.W. and E.C. acknowledge the
Scientific Computer Section at ICTP-Triste, Italy, for assistance.

\newpage

\section*{Figure captions}

\begin{itemize}
\item {\bf Fig.1}: The different types of fractal aggregates investigated
using optical diffraction theory.
Diffusion limited agregate with 2000 particles.
Laplacian fractal composed of 2500 particles.
Poisson fractal (1300 particles) with structural transition.
Biharmonic fractal (3000 particles) with structural transition.
In the later the circle locates the transition point at 60$\%$ far from the
center.

\item {\bf Fig.2}: Central profile ($k_{y}=0$) of the diffraction spectrum
for a Biharmonic fractal with 1500 particles.

\item {\bf Fig.3}: Structure factor of a typical Biharmonic fractal as in
Fig.2.

\item {\bf Fig.4}: Log-log plot of the angle-averaged intensity $<I(k)>$
against $ k$ for diffusion limited (full squares) and Laplacian (open
squares) agregates with 2000 and 2500 particles, respectively.

\item {\bf Fig.5}: Log-log plot of the angle-averaged $<I(k)>$ against
$ k$ for a Poisson agregate as in Fig.1.
Full squares: before the structural transition (650 particles).
Open squares: after the structural transition.

\item {\bf Fig.6}: Log-log plot of the angle-averaged $<I(k)>$ against
$ k$ for a Biharmonic fractal as in Fig.1.
Full squares: before the structural transition (1500 particles).
Open squares: after the structural transition.

\end{itemize}

\end{document}